\documentclass[12pt]{article}
\usepackage{graphicx}
\begin{document} 

\begin{center}
\textbf{Approximately Independent Features of Languages}

\bigskip
Eric W. Holman

\medskip
\textit{Department of Psychology, University of California}\newline
\textit{Los Angeles, California 90095-1563, U. S. A.}\newline
\textit{holman@psych.ucla.edu}
\end{center}

\bigskip
\noindent
To facilitate the testing of models for the evolution of languages, the present note offers a set of linguistic features that are approximately independent of each other.  To find these features, the adjusted Rand index ($R^\prime$) is used to estimate the degree of pairwise relationship among 130 linguistic features in a large published database.  Many of the $R^\prime$ values prove to be near 0, as predicted for independent features, and a subset of 47 features is found with an average $R^\prime$ of -0.0001.  These 47 features are recommended for use in statistical tests that require independent units of analysis.


\bigskip
\noindent
\textbf{1.  Introduction}

\medskip
\noindent
Several stochastic models have recently been developed to simulate the evolution of languages.  The Schulze model [15] and the Viviane model [11,12] in particular have gained empirical realism by representing languages in terms of their features and by successfully approximating the distribution of number of speakers per language.  More specific empirical tests of such models are hampered, however, by the difficulty of dividing linguistic data into the independent units of analysis necessary for conventional statistical inference [3,13].  Separate language families may be phylogenetically independent in the sense of not being descendants of a known common ancestor, but even languages in different families may be related by diffusion, which can extend to continental distances [3,7,10].  Continent-sized areas have in fact been successfully used as units of analysis [3,4], but tests with only five or six such units require very large effects to demonstrate statistically significant results.

An alternative solution would be to use language features, rather than language families or areas, as units of analysis.  A large body of research has explored relationships among linguistic features, starting with Greenberg's famous study of correlations involving the sequential ordering of linguistic elements [5], but little attention has been devoted to the complementary task of finding features that are independent of each other.  The number of independent features has been informally estimated in the context of language learning as about 30 features [2] or about 40 to 50 features [18], numbers large enough to allow tests with adequate statistical power if the independent features themselves can be identified.  The present note addresses the latter task by identifying a set of 47 approximately independent linguistic features.

\bigskip
\noindent
\textbf{2.  Materials and Methods}

\medskip
\noindent
The data are obtained from \textit{The World Atlas of Language Structures} [6], henceforth WALS.  WALS contains 138 maps showing the distribution of different phonological, grammatical, and lexical features, each of which has from two to nine discrete values.  These maps refer to a total of 2560 languages, although few features are attested for more than 1000 languages.  The present study excludes the four features with redundant data, and also the four features referring to color terms, which are attested for a sample of languages that does not overlap enough with the rest to allow reliable comparisons; thus, 130 features are compared.  The analyses are based on 2488 languages, excluding pidgins, creoles, and sign languages.

Relationships among features are typically studied in two-way contingency tables, and numerous summary measures of relationship have been proposed for such tables [1]. Probably the most commonly used measure is the adjusted Rand index, which can be defined as follows in the present context.  Let two features be given, and let $M$ be the number of pairs of languages for which both features are attested in both languages.  Let $A$ be the number of pairs for which both features have the same value in both languages, plus the number of pairs for which both features have different values in the two languages; $A$ is thus the number of language pairs for which the two features agree on the similarity of the languages.  The original Rand index, called $R$, is defined as $A/M$, the proportion of pairs for which the two features agree [14].  $R$ takes the value 1 if the two features always agree, but it is also positive even if the features agree no more than expected by chance.  To correct for chance agreement, the adjusted index $R^\prime$ is defined as $[R-E(R)]/[1-E(R)]$, where $E(R)$ is the expected value of $R$ if the features are independent of each other [9].  $R^\prime$ is still equal to 1 if the features agree perfectly, but it also has expected value 0 if the features are independent.  Negative $R^\prime$ values can occur by chance, but nonrandom relationships between features produce positive $R^\prime$ values on average.  Although the original definition of $R^\prime$ used a large-sample approximation for $E(R)$, the current definition uses an exact formula applicable to any sample size [8]. 

\bigskip
\noindent
\textbf{3.  Results}

\begin{figure}
\includegraphics[angle=-90,scale=0.5]{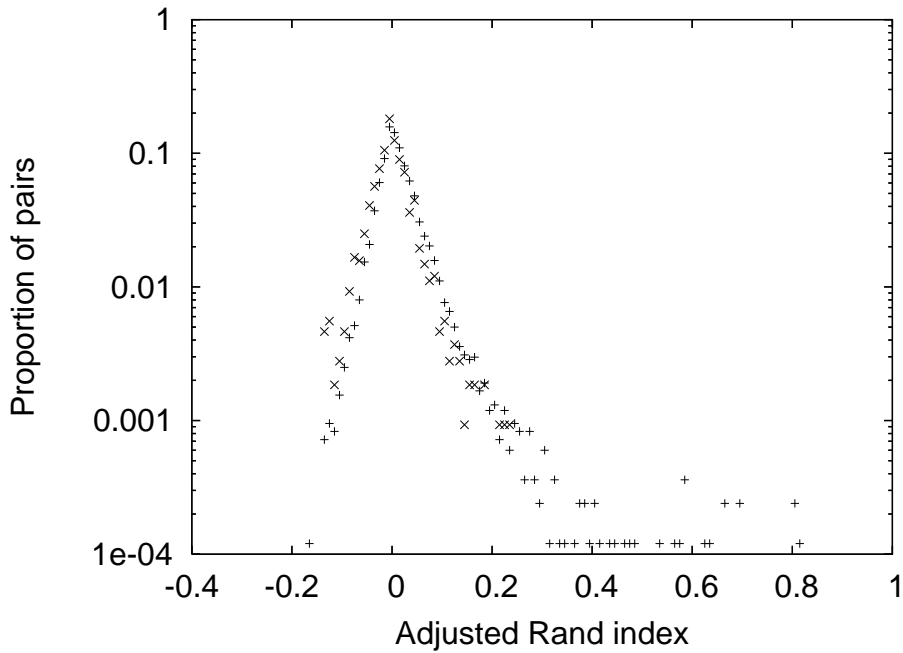}
\caption{Relative frequency distribution of the adjusted Rand index ($R^\prime$), for pairs of all 130 features ($+$), and for pairs of 47 approximately independent features ($\times$).}
\end{figure}
\bigskip

\medskip
\noindent
A matrix of $R^\prime$ for each pair of 130 features contains $130\times129/2$ or 8385 entries.  As a summary of this information, Fig. 1 gives the relative frequency distribution of the $R^\prime$ values in intervals of length 0.01, with proportions scaled logarithmically to encompass the wide range observed.  Each $+$ symbol in the figure represents the proportion of the 8385 values that fall in a particular interval. 

The average $R^\prime$ in the data is 0.0161.  The long upper tail of the distribution indicates that some features are closely related to each other, but the fact that most $R^\prime$ values are small suggests that many of the features may be independent or approximately independent of each other.  A sequential procedure was used to identify a subset of approximately independent features.  Starting with all 130 features, the feature with the highest average $R^\prime$ with the other features was excluded and the average $R^\prime$ among the remaining features was recalculated.  This process was repeated until the average $R^\prime$ among the remaining features was no longer positive.  At this point, there were 47 features left and the average $R^\prime$ among them was --0.0001. 

An immediate question is whether a better procedure would find more than 47 independent features.  As an answer, two types of alternative procedure were explored for sequentially excluding features.  The first alternative used a different criterion for choosing the features to be excluded.  At each step, the pair of features with the highest $R^\prime$ was found, and the member of that pair with the highest average $R^\prime$ with the other remaining features was excluded.  This criterion produced only 11 independent features when used on each step, and 34 independent features when alternated with the original criterion on successive steps.  The second alternative was a type of simulated annealing.  At each step, the features were ranked according to their average $R^\prime$ with the other remaining features, and the probability of excluding each feature was determined by its rank according to a geometric distribution.  The entire process was repeated at least 1000 times for each of six parameter values of the geometric distribution, ranging from 0.2 to 0.8.  Along with many sets of fewer features, several different sets of 47 features were found, but they differed in only a few features from the set originally found, and their average $R^\prime$ was slightly higher.  The original 47 features were therefore retained.

The $×$ symbols in Figure 1 show the distribution of $R^\prime$ values among the 47 features.  This distribution is more symmetric than the one for all 130 features.  There are relatively more negative values and fewer positive ones.  In particular, the long upper tail is gone, and no $R^\prime$ values are above 0.24.

Table 1 lists the 47 approximately independent features.  The first column gives the number of the map for the feature among the 138 maps in WALS, and the second column quotes the summary description of the feature in WALS.  Because of the well known correlations among features involving word order [5], only one feature in the table refers to word order (Feature 92).  Otherwise, the 47 features are fairly evenly distributed across WALS.

\bigskip
\medskip
\noindent

\noindent
\newpage
Table 1.  Approximately independent features.
\medskip

\noindent
Feature no. -- Description\\ 
\medskip
\noindent
1 -- Consonant Inventories \\ 
5 -- Voicing and Gaps in Plosive Systems \\  
8 -- Lateral Consonants \\  
9 -- The Velar Nasal \\  
11 -- Front Rounded Vowels \\  
12 -- Syllable Structure \\  
14 -- Fixed Stress Locations \\  
19 -- Presence of Uncommon Consonants \\  
20 -- Fusion of Selected Inflectional Formatives \\  
22 -- Inflectional Synthesis of the Verb \\  
27 -- Reduplication \\  
30 -- Number of Genders \\  
33 -- Coding of Nominal Plurality \\  
35 -- Plurality in Independent Personal Pronouns \\  
37 -- Definite Articles \\  
41 -- Distance Contrasts in Demonstratives \\  
43 -- Third Person Pronouns and Demonstratives \\  
45 -- Politeness Distinctions in Pronouns \\  
47 -- Intensifiers and Reflexive Pronouns \\  
48 -- Person Marking on Adpositions \\  
52 -- Comitatives and Instrumentals \\  
55 -- Numeral Classifiers \\  
58 -- Obligatory Possessive Inflection \\  
60 -- Genitives, Adjectives and Relative Clauses \\  
61 -- Adjectives without Nouns \\  
65 -- Perfective/Imperfective Aspect \\  
72 -- Imperative-Hortative Systems \\  
73 -- The Optative \\  
74 -- Situational Possibility \\  
79 -- Suppletion According to Tense and Aspect \\  
92 -- Position of Polar Question Particles \\  
99 -- Alignment of Case Marking of Pronouns \\  
100 -- Alignment of Verbal Person Marking \\  
104 -- Order of Person Markers on the Verb \\  
108 -- Antipassive Constructions \\  
109 -- Applicative Constructions \\  
111 -- Nonperiphrastic Causative Constructions \\  
112 -- Negative Morphemes \\  
114 -- Subtypes of Asymmetric Standard Negation \\  
115 -- Negative Indefinite Pronouns and Predicate Negation \\  
120 -- Zero Copula for Predicate Nominals \\  
123 -- Relativization on Obliques \\  
128 -- Utterance Complement Clauses \\  
129 -- Hand and Arm \\  
130 -- Finger and Hand \\  
131 -- Numeral Bases \\  
137 -- N-M Pronouns \\  

\bigskip
\bigskip
\noindent
\textbf{4.  Discussion}

\medskip
\noindent
Statistical inference based on these 47 features as units of analysis requires the assumption that the features are a random sample from a population of independent features.  Results are then generalizable to other features in the same population, which might be sampled in future compilations similar to WALS.  No statistical assumptions are made about the languages themselves, which may be related by inheritance, diffusion, or both.  Strictly speaking, however, results are applicable only to the languages in WALS and are not necessarily generalizable to other languages.

A more important limitation is that only some sorts of questions can be tested with independent features as units of analysis.  Obviously excluded are questions about correlations among features, such as those related to word order [5].  Also excluded are comparisons between values of a single feature, such as whether subject-verb-object order is more frequent than subject-object-verb [3].  The questions that can be tested refer to properties that apply to all features and can be estimated for individual features.  These are in fact the sorts of properties that are typically embodied in models of language evolution.

An example of such a property is the stability of features through time.  The simplest assumption for a model is that all features are equally stable.  A method has recently been developed to estimate empirically the stability of each of the 134 nonredundant WALS features [16].  When stability was estimated separately for languages spoken in the Eastern and Western hemispheres, the rank correlation across features was 0.51, suggesting that stability is a worldwide property that differs among features, but the correlation was not tested statistically because the features are not independent.  Recalculated across the 43 features in Table 1 that are sufficiently attested in both hemispheres, the correlation proves to be 0.43, which is significantly positive, $p < 0.01$, confirming the earlier suggestion.

Another testable question is whether languages with many speakers evolve more slowly than those with few speakers.  In the Viviane model, evolutionary rate is inversely proportional to population [11,12]; and in the Schulze model, the inverse effect of population ranges from negligible to substantial depending upon parameter settings [17].  A recent analysis of the WALS data found a small effect of population; estimated separately in each of the 47 independent features, the correlation between population and instability turned out to be significantly negative [17].  

\bigskip
\noindent
\textbf{Acknowledgments}

\medskip
\noindent
This paper grew out of collaborative research with Cecil H. Brown, Dietrich Stauffer, and S{\o}ren Wichmann, to whom I am grateful for many stimulating discussions.  I also particularly thank Reinhard K\"ohler, Dietrich Stauffer, and S{\o}ren Wichmann for helpful comments on previous drafts of the paper, and Hans-J\"org Bibiko for making the WALS data available in suitable format.

\bigskip
\noindent
\textbf{References}

\medskip
\noindent
[1]  A. N. Albatineh, M. Niewiadomska-Bugaj, and D. Hihalko.  2006.  On Similarity Indices and Correction for Chance Agreement.  \textit{Journal of Classification}, 23, 301--313.\newline
[2]  T. Briscoe.  2000.  Grammatical Acquisition: Inductive Bias and Coevolution of Language and the Language Acquisition Device.  \textit{Language}, 76, 245—296.\newline
[3]  M. S. Dryer.  1989.  Large Linguistic Areas and Language Sampling.  \textit{Studies in Language}, 13, 257--292.\newline
[4]  M. S. Dryer.  1992.  The Greenbergian Word Order Correlations. \textit{Language}, 68, 81--138.\newline
[5]  J. H. Greenberg.  1963.  Some Universals of Grammar with Particular Reference to the Order of Meaningful Elements.  In J. H. Greenberg, ed., \textit{Universals of Language}, MIT Press, Cambridge, MA, 73--113.\newline
[6]  M. Haspelmath, M. Dryer, D. Gil, and B. Comrie, eds.  2005.  \textit{The World Atlas of Language Structures}, Oxford University Press, Oxford.\newline
[7]  E. W. Holman, C. Schulze, D. Stauffer, and S. Wichmann.  2007.  On the Relation between Structural Diversity and Geographical Distance among Languages: Observations and Computer Simulations.  \textit{Linguistic Typology}, 11, 395--423.\newline
[8]  L. Hubert and P. Arabie.  1985.  Comparing Partitions.  \textit{Journal of Classification}, 2, 193--218.\newline
[9]  L. C. Morey and A. Agresti.  1984.  The Measurement of Classification Agreement: an Adjustment of the Rand Statistic for Chance Agreement.  \textit{Educational and Psychological Measurement}, 44, 33--37.\newline
[10]  J. Nichols.  1992.  \textit{Linguistic Diversity in Space and Time}, University of Chicago Press, Chicago.\newline
[11]  V. M. de Olveira, M. A. F. Gomes, and I. R. Tsang.  2006.  Theoretical Model for the Evolution of Linguistic Diversity.  \textit{Physica A}, 361, 361--370.\newline
[12]  P. M. C. de Olveira, D. Stauffer, F. S. W. Lima, A. O. Sousa, C. Schulze, and S. Moss de Olveira. Bit-Strings and Other Modifications of Viviane Model for Language Competition, \textit{Physica A}, 376, 609--616.\newline
[13]  R. D. Perkins.  1989.  Statistical Techniques for Determining Language Sample Size.  \textit{Studies in Language}, 13, 293--315.\newline
[14]  W. M. Rand.  1971.  Objective Criteria for the Evaluation of Clustering Methods.  \textit{Journal of the American Statistical Association}, 66, 846--850.\newline
[15]  C. Schulze, D. Stauffer, and S. Wichmann.  2008.  Birth, Survival and Death of Languages by Monte Carlo Simulation.  \textit{Communications in Computational Physics}, 3, 271--294.\newline
[16]  S. Wichmann and E. W. Holman.  2007.  Assessing Temporal Stability for Linguistic Typological Features.  Manuscript under review.  Available at: http://email.eva.mpg.de/\~{}wichmann/WichmannHolmanIniSubmit.pdf\newline
[17]  S. Wichmann, D. Stauffer, C. Schulze, and E. W. Holman.  2007.  Do Language Change Rates Depend on Population Size?  Manuscript under review.  Available at arXiv:0706.1842\newline
[18]  C. Yang.  2006.  \textit{The Infinite Gift: How Children Learn and Unlearn the Languages of the World}, Scribner, New York.
 
\end{document}